# Gaussian as test functions in Operator Valued Distribution formulation of QED

**Hasimbola Damo Emile Randriamisy[1], Raoelina Andriambololona[2], Hanitriarivo Rakotoson[3]**
**Ravo Tokiniaina Ranaivoson[4], Roland Raboanary[5]**

*Computer Science and Theoretical Physics Department, Institut National des Sciences et Techniques Nucléaires (INSTN-Madagascar), Antananarivo, Madagascar*[1, 2, 3, 4]
*Mention Physique et Applications, Faculté des Sciences – University of Antananarivo*[3, 5]

*hasinadamo@gmail.com (Hasimbola Damo Emile Randriamisy), raoelina.andriambololona@gmail.com(Raoelina Andriambololona)*

*infotsara@gmail.com(Hanitriarivo Rakotoson) , tokhiniaina@gmail.com(Ravo Tokiniaina Ranaivoson),*

*r_raboanary@yahoo.fr (Roland Raboanary)*

**Abstract:** As shown by Epstein and Glaser, the operator valued distribution (OPVD) formalism permits to obtain a non-standard regularization scheme which leads to a divergences-free quantum field theory. We show, with the example of a scalar quantum electrodynamics theory, that Gaussian functions may be used as test functions in this approach. After a short recall about the OPVD formalism in $3 + 1$-dimensions, Gaussian functions and Harmonic Hermite-Gaussian functions are used as test functions. The vacuum fluctuation, Feynman propagators and a study about loop convergence with the example of the tadpole diagram are given. The approach is extended to Quantum Electrodynamics. Calculations concerning triangle anomaly and Ward-Takahashi identity are performed in the framework of the method.

## 1. Introduction

Divergences appear when calculating S-matrix in Quantum Theory [1-3]. Many approaches have been considered to deal with this problem.
The standard renormalization techniques based on the dimensional regularization and Cut-Off method has been used [1-2-4]. Operator Valued Distribution approach (OPVD) developed by Epstein and Glaser [5] based on the mathematical theory of distribution is considered in our paper.
In this work, we denote the quadri-momentum by $\bar{p} = p^\mu e_\mu \; for \; \mu = 0,1,2,3$ where $\{e_\mu\}$ is a basis in the Minkowski space and the tri-momentum by $\vec{p} = p^i \vec{e}_i \; for \; i = 1,2,3$ in which $\{\vec{e}_i\}$ is the 3-dimensional Euclidian basis. The corresponding scalar products are $\bar{p}.\bar{x} = p^\mu x_\mu$ and $\vec{p}.\vec{x} = p^i x_i$.
In the Minkowski space with signature $(1, 3)$, the field is defined by

$$\boldsymbol{\phi}(\bar{x}) = \langle \mathbb{T}_{\bar{x}} \boldsymbol{\Phi}, g \rangle \qquad (1.1)$$

in which
- $g$ is a test function
- $\boldsymbol{\Phi}$ is an OPVD [5-9]: it is an operator on a Fock space and a distribution acting on the set $\{g\}$ of test functions.
- $\mathbb{T}_{\bar{x}}$ is the translation operator defined by the relations
$$\begin{cases} \mathbb{T}_{\bar{x}} g(\bar{y}) = g(\bar{y} + \bar{x}) \\ \langle \mathbb{T}_{\bar{x}} \boldsymbol{\Phi}, g \rangle = \langle \boldsymbol{\Phi}, \mathbb{T}_{-\bar{x}} g \rangle \end{cases}$$

The explicit expression of the field is [7]

$$\boldsymbol{\phi}(\vec{x},t) = \int \frac{d^3\vec{p}}{(\sqrt{2\pi})^3} \frac{1}{\sqrt{2\omega}} [\boldsymbol{a}(\vec{p})e^{-i\bar{p}.\bar{x}} + \boldsymbol{b}^\dagger(\vec{p})e^{i\bar{p}.\bar{x}}] \tilde{g}(\omega,\vec{p}) \tag{1.2}$$

in which

- $\boldsymbol{a}(\vec{p})$ and $\boldsymbol{b}^\dagger(\vec{p})$ are respectively particle annihilation and antiparticle creation operators satisfying the commutation relations

$$\begin{cases} [\boldsymbol{a}(\vec{p}), \boldsymbol{b}^\dagger(\vec{q})] = [\boldsymbol{a}(\vec{p}), \boldsymbol{b}(\vec{q})] = 0 \\ [\boldsymbol{a}(\vec{p}), \boldsymbol{a}^\dagger(\vec{q})] = \delta^3(\vec{p} - \vec{q}) \\ [\boldsymbol{b}(\vec{p}), \boldsymbol{b}^\dagger(\vec{q})] = \delta^3(\vec{p} - \vec{q}) \end{cases} \tag{1.3}$$

- $\omega = p_0 = \sqrt{\vec{p}^2 + m^2}$ is the energy of a free particle, $m$ being its mass.
- $\tilde{g}$ is the Fourier transform of the test function $g$

$$\tilde{g}(\bar{p}) = \frac{1}{(\sqrt{2\pi})^4} \int g(\bar{y}) e^{-i\bar{p}\bar{y}} d^4\bar{y} \tag{1.4}$$

The Klein-Gordon equation

$$(\partial_\mu \partial^\mu - m^2)\boldsymbol{\phi}(\bar{x}) = 0 \tag{1.5}$$

for $\mu = 0, \ldots, 3$ is satisfied. We study the cases in which the test functions are Gaussian functions namely Harmonic Hermite-Gaussian functions introduced and studied in the references [10-12]. If we denote the latter $g_n(x)$ and their Fourier transforms $\tilde{g}_n(p)$ their expressions are

$$\begin{cases} g_n(\bar{x}) = Q_n(\bar{x}) e^{-\mathcal{B}_{\mu\nu}(x^\mu - X^\mu)(x^\nu - X^\nu) - iP_\mu x^\mu} \\ \tilde{g}_n(\bar{p}) = \mathcal{R}_n(\bar{p}) e^{-\mathcal{A}^{\mu\nu}(p_\mu - P_\mu)(p_\nu - P_\nu) + iX^\mu(p_\mu - P_\mu)} \end{cases} \tag{1.6}$$

in which $Q_n(\bar{x})$ and $\mathcal{R}_n(\bar{p})$ are polynomials of degree $n$ respectively with variables $x^\mu$ and variables $p_\mu$, $\mu = 0,1,2,3$). $X^\mu$ and $P_\mu$ are respectively the mean values of the variables $x^\mu$ and $p_\mu$

$$\begin{cases} \int |g_n(\bar{x})|^2 d^4\bar{x} = 1 & \int x^\mu |g_n(\bar{x})|^2 d^4\bar{x} = X^\mu \\ \int |\tilde{g}_n(\bar{p})|^2 d^4\bar{p} = 1 & \int p_\mu |\tilde{g}_n(\bar{p})|^2 d^4\bar{p} = P_\mu \end{cases} \tag{1.7}$$

$\mathcal{A}^{\mu\nu}$ and $\mathcal{B}_{\mu\nu}$ are the components of second order tensors representing the coordinate and momentum dispersion-codispersion operators corresponding to the ground state ($n=0$) [11-13].

$$\begin{cases} \int (x^\mu - X^\mu)(x^\nu - X^\nu) |g_0(\bar{x})|^2 d^4\bar{x} = \mathcal{A}^{\mu\nu} \\ \int (p_\mu - P_\mu)(p_\nu - P_\nu) |\tilde{g}_0(\bar{p})|^2 d^4\bar{p} = \mathcal{B}_{\mu\nu} \\ \mathcal{A}^{\mu\rho} \mathcal{B}_{\rho\nu} = \frac{1}{4} \delta^\mu_\nu \end{cases} \tag{1.8}$$

The polynomials $Q_n(\bar{x})$ and $\mathcal{R}_n(\bar{p})$ in the relation (1.6) can be expressed in terms of Hermite polynomials [11].

For the case $\mathcal{A}^{\mu\nu} = \mathcal{B}_{\mu\nu} = 0$ for $\mu \neq \nu$, we have uncorrelated Harmonic Hermite Gaussian functions

$$\begin{cases} Q_n(\bar{x}) = \prod_{\mu=0}^{3} \dfrac{H_{n^\mu}(\frac{x^\mu - X^\mu}{\sqrt{2\mathcal{A}^{\mu\mu}}})}{\sqrt{2^{n^\mu} n^\mu! (\sqrt{2\pi})^{3+1} \sqrt{\mathcal{A}^{\mu\mu}}}} \\ \mathcal{R}_n(\bar{p}) = \prod_{\mu=0}^{3} \dfrac{H_{n^\mu}(\frac{p_\mu - P_\mu}{\sqrt{2\mathcal{B}_{\mu\mu}}})}{\sqrt{2^{n^\mu} n^\mu! (\sqrt{2\pi})^{3+1} \sqrt{\mathcal{B}_{\mu\mu}}}} \end{cases} \quad (1.9)$$

in which $H_{n^\mu}$ is the Hermite polynomials of degree $n^\mu$. For the ground state, $n^\mu = 0\ for\ \mu = 0,1,2,3$, the relations (1.6) give

$$\begin{cases} g_0(\bar{x}) = \dfrac{e^{-\mathcal{B}_{\mu\nu}(x^\mu - X^\mu)(x^\nu - X^\nu) - iP_\mu x^\mu}}{\sqrt{(\sqrt{2\pi})^{3+1} \prod_{\mu=0}^{3}(\sqrt{\mathcal{A}^{\mu\mu}})}} \\ \tilde{g}_0(\bar{p}) = \dfrac{e^{-\mathcal{A}^{\mu\nu}(p_\mu - P_\mu)(p_\nu - P_\nu) + iX^\mu(p_\mu - P_\mu)}}{\sqrt{(\sqrt{2\pi})^{3+1} \prod_{\mu=0}^{3}(\sqrt{\mathcal{B}_{\mu\mu}})}} \end{cases} \quad (1.10)$$

## 2. Gaussian functions as test functions

In this section, we consider the case in which the test function is the Gaussian function $\tilde{g}_0(\bar{p})$ (1.10). For sake of simplicity we choose $X^\mu = 0$. This approach is suggested by the fact that Gaussian functions correspond to the normal probability distribution and appear in the modeling of many phenomena.

### 2.1. Calculation of vacuum fluctuation

Let us consider the vacuum fluctuation

$$\vartheta(\bar{x}, \bar{y}) = \langle 0|\boldsymbol{\phi}(\bar{x})\boldsymbol{\phi}^\dagger(\bar{y})|0\rangle \quad (2.1)$$

$|0\rangle$ being the vacuum state and

$$\boldsymbol{\phi}^\dagger(\bar{y}) = \int \frac{d^3\vec{p}}{(\sqrt{2\pi})^3} \frac{1}{\sqrt{2\omega}} [\boldsymbol{a}^\dagger(\vec{p}) e^{i\vec{p}\cdot\bar{y}} + \boldsymbol{b}(\vec{p}) e^{-i\vec{p}\cdot\bar{y}}] \tilde{g}_0(\omega, \vec{p}) \quad (2.2)$$

We obtain

$$\vartheta(\bar{x}, \bar{y}) = \int \frac{d^3\vec{p}}{(\sqrt{2\pi})^3} \frac{d^3\vec{q}}{(\sqrt{2\pi})^3} \frac{1}{2\omega} \langle 0|\boldsymbol{a}(\vec{p})\boldsymbol{a}^\dagger(\vec{q})|0\rangle e^{-i\vec{p}\cdot(\bar{x}-\bar{y})} \tilde{g}_0(\omega, \vec{p}) \tilde{g}_0(\omega', \vec{q}) \quad (2.3)$$

$$= \int \frac{d^3\vec{p}}{(2\pi)^3} \frac{1}{2\omega} d^3\vec{q}\, \delta^3(\vec{p} - \vec{q}) e^{-i\vec{p}\cdot(\bar{x}-\bar{y})} \tilde{g}_0(\omega, \vec{p}) \tilde{g}_0(\omega', \vec{q}) \quad (2.4)$$

The integration of (2.4) over $\vec{q}$ gives

$$\vartheta(\bar{x},\bar{y}) = \int \frac{d^3\vec{p}}{2\omega(2\pi)^3} e^{-i\bar{p}.(\bar{x}-\bar{y})} [\tilde{g}_0(\omega,\vec{p})]^2 \qquad (2.5)$$

This result (2.5) differs from the classical one by the regularizing factor $[\tilde{g}_0(\omega,\vec{p})]^2$.
Using the expression of $\tilde{g}_0$ in (1.10) and taking the case $p_0 = \omega$ we have explicitly

$$\vartheta(\bar{x},\bar{y}) = \int \frac{d^3\vec{p}}{2\omega(2\pi)^3} e^{-i\bar{p}.(\bar{x}-\bar{y})} \frac{e^{-2\mathcal{A}^{\mu\nu}(p_\mu-P_\mu)(p_\nu-P_\nu)}}{\left(\sqrt{2\pi}\right)^3 \prod_{\mu=0}^{3}(\sqrt{\mathcal{B}_{\mu\mu}})} \qquad (2.6)$$

**2.2. Calculation of Feynman propagator**

The Feynman propagator in the configuration space is

$$\Delta_F(\bar{x}-\bar{y}) = -i\langle 0|T\{\boldsymbol{\phi}(\bar{x})\boldsymbol{\phi}^\dagger(\bar{y})\}|0\rangle \qquad (2.7)$$

in which $T$ is the chronological product [1].

$$T\{\boldsymbol{\phi}(\bar{x})\boldsymbol{\phi}^\dagger(\bar{y})\} = \begin{cases} \boldsymbol{\phi}(\bar{x})\boldsymbol{\phi}^\dagger(\bar{y}) \text{ if } x^0 \geq y^0 \\ \boldsymbol{\phi}^\dagger(\bar{y})\boldsymbol{\phi}(\bar{x}) \text{ if } y^0 \geq x^0 \end{cases} \qquad (2.8)$$

then

$$\langle 0|T\{\boldsymbol{\phi}(\bar{x})\boldsymbol{\phi}^\dagger(\bar{y})\}|0\rangle = \theta(t)\langle 0|\boldsymbol{\phi}(\bar{x})\boldsymbol{\phi}^\dagger(\bar{y})|0\rangle + \theta(-t)\langle 0|\boldsymbol{\phi}^\dagger(\bar{y})\boldsymbol{\phi}(\bar{x})|0\rangle \qquad (2.9)$$

in which $t = x^0 - y^0$ in natural units ($c = 1$) and $\theta$ is the Heaviside function.
Taking into account the expression (1.2) an (2.2) of fields, we obtain

$$\langle 0|T\{\boldsymbol{\phi}(\bar{x})\boldsymbol{\phi}^\dagger(\bar{y})\}|0\rangle = \int \frac{d^4\bar{p}}{(2\pi)^4} \frac{i[\tilde{g}_0(p)]^2}{[(p^0)^2 - (\omega_p)^2 + i\varepsilon]} e^{-i\bar{p}.(\bar{x}-\bar{y})} \qquad (2.10)$$

$$= \int \frac{d^4\bar{p}}{(2\pi)^4} \frac{i[\tilde{g}_0(p)]^2}{[(p^0)^2 - \vec{p}^2 - m^2 + i\varepsilon]} e^{-i\bar{p}.(\bar{x}-\bar{y})} \qquad (2.11)$$

The Feynman propagator in the configuration space is

$$\Delta_F(\bar{x}-\bar{y}) = \int \frac{d^4\bar{p}}{(2\pi)^4} \frac{e^{-i\bar{p}.(\bar{x}-\bar{y})}}{[\bar{p}^2 - m^2 + i\varepsilon]} \frac{e^{-2\mathcal{A}^{\mu\nu}(p_\mu-P_\mu)(p_\nu-P_\nu)}}{\left(\sqrt{2\pi}\right)^4 \prod_{\mu=0}^{3}(\sqrt{\mathcal{B}_{\mu\mu}})} \qquad (2.12)$$

The Feynman propagator in momentum space is

$$\tilde{\Delta}_F(\bar{p}) = \frac{1}{[\bar{p}^2 - m^2 + i\varepsilon]} \frac{e^{-2\mathcal{A}^{\mu\nu}(p_\mu-P_\mu)(p_\nu-P_\nu)}}{\left(\sqrt{2\pi}\right)^4 \prod_{\mu=0}^{3}(\sqrt{\mathcal{B}_{\mu\mu}})} \qquad (2.13)$$

## 3. Harmonic Hermite-Gaussian functions as test functions

We can also use as test functions the Harmonic Hermite-Gaussian functions defined in the relation (1.6). The vacuum fluctuation and the Feynman propagators (with $X^\mu = 0$ and $p^0 = \omega$) are in this case

$$\langle 0|\boldsymbol{\phi}(\bar{x})\boldsymbol{\phi}^\dagger(\bar{y})|0\rangle = \int \frac{d^3\vec{p}}{2\omega(2\pi)^3} e^{-i\bar{p}.(\bar{x}-\bar{y})} \mathcal{R}_n(\omega,\vec{p}) \frac{e^{-2\mathcal{A}^{\mu\nu}(p_\mu - P_\mu)(p_\nu - P_\nu)}}{(\sqrt{2\pi})^4 \prod_{\mu=0}^3 (\sqrt{\mathcal{B}_{\mu\mu}})} \tag{3.1}$$

$$\Delta_F(\bar{x}-\bar{y}) = \int \frac{d^4\bar{p}}{(2\pi)^4} \frac{1}{[\bar{p}^2 - m^2 + i\varepsilon]} e^{-i\bar{p}.(\bar{x}-\bar{y})} \mathcal{R}_n(\bar{p}) \frac{e^{-2\mathcal{A}^{\mu\nu}(p_\mu - P_\mu)(p_\nu - P_\nu)}}{(\sqrt{2\pi})^4 \prod_{\mu=0}^3 (\sqrt{\mathcal{B}_{\mu\mu}})} \tag{3.2}$$

$$\tilde{\Delta}_F(\bar{p}) = \frac{1}{[\bar{p}^2 - m^2 + i\varepsilon]} \mathcal{R}_n(\bar{p}) \frac{e^{-2\mathcal{A}^{\mu\nu}(p_\mu - P_\mu)(p_\nu - P_\nu)}}{(\sqrt{2\pi})^4 \prod_{\mu=0}^3 (\sqrt{\mathcal{B}_{\mu\mu}})} \tag{3.3}$$

The difference between the expressions (2.1), (2,3) and (3.1), (3.2), (3.3) is the presence of the polynomials $\mathcal{R}_n(\bar{p})$ which is reduced to a product of Hermite polynomials in the case of uncorrelated variables.

## 4. Application to one loop scalar Quantum Electrodynamics diagram

One loop scalar tadpole diagram contributes to the two point correlation function. We calculate the associated amplitude since it is simple and the regularization factor appears.

### 4.1. Divergence in the framework of standard formulation

We apply the above results for the calculation of more complicated entities considered in the scalar field theory. Let us, for instance, perform the calculation corresponding to the tadpole diagram which appear in the calculation of photon propagator in conventional scalar QED [1]

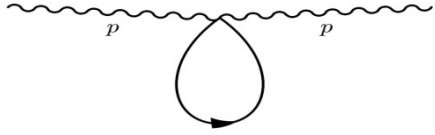

*Fig 1.One loop diagram*

In old formulation, the amplitude of this tadpole diagram is given by

$$\mathcal{M}_{\mu\nu} = 2e^2 g_{\mu\nu} \int \frac{d^4\bar{k}}{(2\pi)^4} \frac{i}{\bar{k}^2 - m^2 + i\varepsilon} \tag{4.1}$$

Introducing quadridimensional spherical coordinate $(k_E, \vartheta, \theta, \varphi)$, (4.1) becomes

$$\mathcal{M}_{\mu\nu} = \frac{e^2 g_{\mu\nu}}{8\pi^4} \int_0^{+\infty} \int_0^\pi \int_0^\pi \int_0^{2\pi} \frac{k_E^3}{k_E^2 + m^2} (\sin\vartheta)^2 (\sin\theta) dk_E d\theta d\vartheta d\varphi$$

$$= \frac{e^2 g_{\mu\nu}}{4\pi^2} \left[ \frac{k_E^2}{2} + \frac{m^2}{2} \ln\left(\frac{k_E^2}{m^2} + 1\right) \right]_0^{+\infty} \tag{4.2}$$

where

$$k_E^2 = k_0^2 + \vec{k}^2 \tag{4.3}$$

The expression (4.2) shows quadratic and logarithmic divergences. In the standard formulation, one deals with this divergence using regularization and renormalization techniques [1, 2].

## 4.2. Calculation in the framework of OPVD formulation using Gaussian functions

Let us now use the Gaussian functions defined in the relation (1.6). The Feynman propagator in momentum space is given by the relation (2.13). The analog of the integral (4.1) is then

$$\mathcal{M}_{\mu\nu} = 2e^2 g_{\mu\nu} \int \frac{d^4 k_E}{(2\pi)^4} \frac{1}{k_E^2 + m^2} [\tilde{g}_0(k_E^2)]^2 \qquad (4.4)$$

As we are mainly interested in the study of convergence, we consider the case

$$\begin{cases} \mathcal{A}^{\mu\nu} = 0 \ if \ \mu \neq \nu \\ \mathcal{A}^{\mu\mu} = \mathcal{A} = \frac{1}{4\mathcal{B}} \\ K_\mu = 0 \end{cases} \qquad (4.5)$$

The integral (3.8) becomes

$$\mathcal{M}_{\mu\nu} = 2e^2 g_{\mu\nu} \int \frac{d^4 k_E}{(2\pi)^4} \frac{1}{k_E^2 + m^2} e^{-\frac{k_E^2}{2\mathcal{B}}} \qquad (4.6)$$

then

$$\mathcal{M}_{\mu\nu} = 4\pi^2 e^2 g_{\mu\nu} \int_0^{+\infty} \frac{k_E^3 e^{-\frac{k_E^2}{2\mathcal{B}}}}{k_E^2 + m^2} dk_E \qquad (4.7)$$

Performing the variable change $u = \frac{k_E}{\sqrt{2\mathcal{B}}}$, we obtain

$$\mathcal{M}_{\mu\nu} = \frac{e^2 g_{\mu\nu} \mathcal{B}}{2\pi^2} \int_0^{+\infty} \frac{u^3 e^{-u^2}}{u^2 + (\frac{m}{\sqrt{\mathcal{B}}})^2} du \qquad (4.8)$$

Following the interpretation in the references [11-13], the parameter $\sqrt{\mathcal{B}}$ is the common value of the uncertainty (statistical standard deviation) on the values of the energy $k_0$ and the components $k_j$ ($j = 1,2,3$) of the spatial momentum of the particle corresponding to the tadpole. In the very large energy approximation, $(\frac{m}{\sqrt{\mathcal{B}}})^2 \to 0$, the integral (4.8) becomes

$$\mathcal{M}_{\mu\nu} = \frac{e^2 \mathcal{B}}{2\pi^2} g_{\mu\nu} \int_0^{+\infty} u e^{-u^2} du = \frac{e^2}{4\pi^2} \mathcal{B} g_{\mu\nu} \qquad (4.9)$$

As expected, we obtain a finite result. This result relates explicitly the existence of loop in the photon propagator of scalar QED to the Heisenberg uncertainty relation. This relation is obvious since the Heisenberg principle is already integrated in the representation of quantum phase space formalism.

# 5. Triangle axial anomaly in QED

Let us use Gaussian test function to calculate diagram in QED which involved current conservation. Axial anomaly can be understood by calculating the two triangles graphs

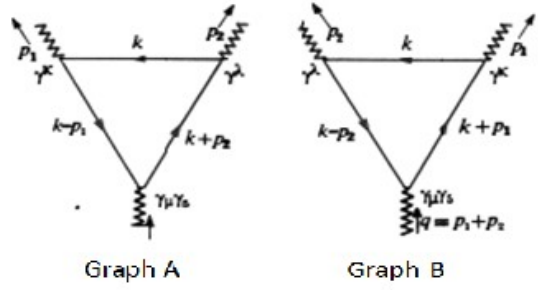

Graph A             Graph B

The amplitude of the two graphs contributing to the triangle anomaly can be written as [2]

$$T_{k,\lambda,\mu}(p_1,p_2) = I^1_{k,\lambda,\mu}(p_1,p_2) + I^2_{\lambda,k,\mu}(p_1,p_2) \tag{5.1}$$

$I^1_{k,\lambda,\mu}(p_1,p_2)$ is the amplitude of the Graph A and $I^1_{k,\lambda,\mu}(p_1,p_2)$ is the amplitude of the Graph B

We will show the conservation of the axial currents at the vertices [2]:

$$(p_1+p_2)^\mu T_{k,\lambda,\mu} = 0 \tag{5.2}$$

Using Feynman rules and after some algebra calculation we obtain for the amplitude

$$(p_1+p_2)^\mu \left(I^1_{k,\lambda,\mu} + I^2_{k,\lambda,\mu}\right) = 4e^2 \epsilon_{\sigma\lambda\delta k} \int \frac{d^4\bar{k}}{(2\pi)^4} \frac{k^\delta}{\bar{k}^2} \{p_1^\sigma \Delta\tilde{g} - p_2^\sigma \Delta\tilde{g}'\} \tag{5.3}$$

where

$$\begin{cases} \Delta\tilde{g} = \tilde{g}\left(\frac{\bar{k}^2}{2\mathcal{B}}\right)\left[\tilde{g}\left(\frac{(\bar{k}-\bar{p}_2)^2}{2\mathcal{B}}\right) - \tilde{g}\left(\frac{(\bar{k}+\bar{p})^2}{2\mathcal{B}}\right)\right] \\ \Delta\tilde{g}' = \tilde{g}\left(\frac{\bar{k}^2}{2\mathcal{B}}\right)\left[\tilde{g}\left(\frac{(\bar{k}+\bar{p}_1)^2}{2\mathcal{B}}\right) - \tilde{g}\left(\frac{(\bar{k}-\bar{p}_1)^2}{2\mathcal{B}}\right)\right] \end{cases} \tag{5.4}$$

The result (5.4) shows that relation (5.2) is satisfied in the limit where $\bar{p}_1^2, \bar{p}_2^2 \ll 2\mathcal{B}$. In this limit too, $\tilde{g}$ depends only on the $k^2$ variable everywhere and we have

$$\begin{cases} \Delta\tilde{g} = \tilde{g}\left(\frac{\bar{k}^2}{2\mathcal{B}}\right)\left[\tilde{g}\left(\frac{\bar{k}^2}{2\mathcal{B}}\right) - \tilde{g}\left(\frac{\bar{k}^2}{2\mathcal{B}}\right)\right] = 0 \\ \Delta\tilde{g}' = \tilde{g}\left(\frac{\bar{k}^2}{2\mathcal{B}}\right)\left[\tilde{g}\left(\frac{\bar{k}^2}{2\mathcal{B}}\right) - \tilde{g}\left(\frac{\bar{k}^2}{2\mathcal{B}}\right)\right] = 0 \end{cases} \tag{5.5}$$

The conservation of the axial current is then verified by our canonical regularized quantum field.
It may be shown that we have the conservation of the axial current in general case (see section 6)
In the following, we will give another Gaussian type function with which we could define a partition of unity [5], which is more generalized.

## 6. Field as OPVD with Gaussian test Function

The amplitude in the OPVD formalism is

$$A = \int_0^\infty dX\, T(X) g(X) \tag{6.1}$$

where $g(X)$ is the Gaussian test function. The calculation will be done in one dimension for sake of simplicity. Four-dimensional extension can be achieved using integration by part.

### 6.1. Gaussian test function as partition of unity

We reduce the test function to a partition of unity following the procedure in [5-7] by the convolution of the Gaussian function of the form

$$g(x) = \begin{cases} \mathcal{N} e^{-x^2} & for\ \|x\| < 1 \\ 0 & elsewhere \end{cases} \tag{6.2}$$

where $\mathcal{N}$ is a normalization factor. The test function as partition of unity will be denoted $f$ and it can be tend to 1 over the whole domain in a given limit.

### 6.2. Construction of the distribution extension in the UV limit

In this section, we use Lagrange formula for the case of large momentum. In quantum field theory, we calculate mostly the amplitude associated to Feynman diagrams. In the OPVD formulation the amplitude could be defined as

$$A = \int_0^\infty dX\, \check{T}^>(X) f(X) \tag{6.3}$$

where $\check{T}^>(X)$ is the extension of the distribution $T(X)$ in the UV limit. This amplitude $\mathcal{A}$ is regularized by the Gaussian test function. $\check{T}^>(X)$ is useful in higher order calculation since it is regular.

### 6.3. Extension of the test function

In order to avoid the pole in the amplitude, we modify the test function by using the Taylor remainder [5]. In the UV limit, the test function is equal to its Taylor remainder

$$g^>(X) = -\frac{X}{k!} \int_1^{\eta^2 G_\alpha(\|X\|)} \frac{dt}{t} (1-t)^k \partial_X^{k+1}\left(X^k f^>(Xt)\right) \tag{6.4}$$

Here, the upper born, $\eta^2 G_\alpha(\|X\|)$, is given by the arbitrary choice of a running support. Following the procedure described in [5], we change $f^>(X)$ for this value and get

$$A = -\int_0^\infty dX\, T(X) \frac{X}{k!} \int_1^{\eta^2 G_\alpha(\|X\|)} \frac{dt}{t} (1-t)^k \partial_X^{k+1}\left(X^k f^>(Xt)\right) \tag{6.5}$$

### 6.4. Extension of the distribution $T(X)$

Integrating by parts (6.5) and performing the limit $f^>(Xt) \to 1$, which is now possible since the distribution is regularized, we obtain the definition of the extension of the distribution in the UV domain [4] [5] [6].

$$\check{T}^>(X) = \frac{(-X)^k}{k!} \partial_X^{k+1}\left(X T(X)\right) \int_1^{\eta^2} \frac{dt}{t} (1-t)^k \tag{6.6}$$

This Lagrange formula gives a divergence-free contribution.

## 7. Ward-Takahashi identity

In order to keep the gauge symmetry of the theory, we calculate the Ward-Takahashi identity in one loop order. This identity is given by [2]

$$\delta \Gamma^\mu = -\frac{\partial}{\partial p} \Sigma(p^2) \Big|_{\gamma p = m} \tag{7.1}$$

where $\Sigma(p^2)$ is the selfenergy amplitude at one loop order and $\delta\Gamma^\mu$ is the vertex amplitude at one loop order. The Ward-Takahashi identity is a powerful tool used in QED to show that the gauge symmetry of the theory remains after the regularization procedure [2]. This was used to show that some of the conventional regularization procedure breaks the gauge symmetry of the Lagrangian. This is the case of dimensional regularization, hard cut-off procedure and so on. Let us calculate the amplitude involved in the Ward-Takahashi identity. Two diagrams contribute to this identity at one-loop order. The first diagram, Graph C, is the self-energy. In the old fashioned formulation of QED, this diagram contribution is UV-divergent. The second diagram, Graph D, is the vertex.

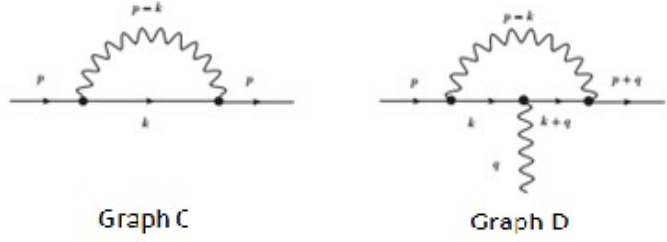

Graph C          Graph D

We use the Lagrange formula described above to calculate the two diagrams amplitude.

### 7.1. One loop Self-energy diagram (Graph C)

Let us calculate the one loop order contribution to propagator, Graph C. The amplitude of this is obtained by using Feynman rules

$$\Sigma(k) = \int \frac{d^4\bar{k}}{(2\pi)^4} \frac{(\gamma^\mu k_\mu + m)}{\bar{k}^2 - m^2} \gamma^\alpha \frac{-i}{(\bar{p}-\bar{k})^2 - \mu^2} \gamma_\alpha g(\bar{k}^2) g\left((\bar{p}-\bar{k})^2\right) \tag{7.2}$$

in which
- $\bar{k}$ is the quadri-momentum
- $g$ is a factor emerging from the field formulation as Operator valued distribution. A photon mass $\mu$ is introduced to avoid IR-divergence [2].
- $\gamma_\alpha$ are the Gamma matrices of Dirac

introducing the Feynman parameterization, the integral becomes

$$\Sigma(\bar{p}) = \int_0^1 dx \int \frac{d^4\bar{k}}{(2\pi)^4} \frac{(2x\gamma\bar{p} - 4m)}{(\bar{k}^2 - M^2(x,\bar{p}^2))^2} f^2\left(\frac{\bar{p}.\bar{x}}{\Lambda^2}\right)^2 f^2\left(\frac{\bar{k}^2}{\Lambda^2}\right) \tag{7.3}$$

where
$$M^2(x, \bar{p}^2) = x(1-x)\bar{p}^2 + xm^2 + (1-x)\mu^2$$

and $\Lambda$ is an arbitrary scale which is related to the dispersion parameter. In order to use the Taylor formula we make the variable change $X = \frac{k^2}{\Lambda^2}$

Then

$$\Sigma(\bar{k}) = \frac{e^2}{16\pi^2} \int_0^1 dx [2x\gamma\bar{p} - 4m] \int_0^{+\infty} \partial_X \frac{X^2}{(X+1)^2} dX \ln\left(\frac{\eta^2}{M^2(x,\bar{p}^2)}\right) f^2\left(X \frac{M^2(z,\bar{p}^2)}{\Lambda^2}\right) \quad (7.4)$$

The integral can be split into the following integrals

$$\begin{cases} I_1 = 1 \int_0^\infty dX \frac{1}{(X+1)} f^2\left(X \frac{M^2(x,\bar{p}^2)}{\Lambda^2}\right) \\ I_2 = -3 \int_0^\infty dX \frac{1}{(X+1)^2} f^2\left(X \frac{M^2(x,\bar{p}^2)}{\Lambda^2}\right) \\ I_3 = 2 \int_0^\infty dX \frac{1}{(X+1)^3} f^2\left(X \frac{M^2(x,\bar{p}^2)}{\Lambda^2}\right) \end{cases} \quad (7.5)$$

The application of the Lagrange formula gives

$$\Sigma(\bar{p}^2) = -\frac{\alpha}{4\pi} \int_0^1 dx [2x\gamma p - 4m] \ln\left[\frac{\eta^2 \Lambda^2}{x(1-x)\bar{p}^2 + xm^2 + (1-x)\mu^2}\right] \quad (7.6)$$

The self-energy diagram contribution we obtained is the same as in the dimensional regularization procedure. For the Ward-Takahashi identity, we need the derivative of the self-energy.

$$\frac{\partial}{\partial \bar{p}} \Sigma(\bar{p}) = -\frac{\alpha}{4\pi} \left[\ln\left(\eta^2 \frac{\Lambda^2}{m^2}\right) - 2\ln\left(\frac{\mu^2}{m^2}\right)\right] + o\left(\frac{m^2}{\Lambda^2}, \frac{\mu^2}{\Lambda^2}\right) \quad (7.7)$$

This result is equivalent to those obtained by conventional calculation up to a constant term which could be introduced in the logarithmic part by some basics properties.

## 7.2. Vertex diagram (Graph D)

The last diagram which contributes to the Ward identity in QED at one loop order is the vertex diagram. It involves the coupling between fermion and photon, with a radiative correction [1] [2].
Let us calculate first the distribution part of the amplitude

$$\delta\Gamma^\mu = \int \frac{d^4\bar{k}}{(2\pi)^4} \frac{1}{(\bar{k}-\bar{p})^2 + \mu^2} \gamma^\nu \frac{(\gamma^\mu k_\mu + m)}{(\bar{k}+\bar{q})^2 - m^2} \gamma^\mu \frac{(\gamma^\mu k_\mu + \gamma^\mu q_\mu + m)}{\bar{k}^2 - m^2} \gamma_\nu \quad (7.8)$$

We use the matrix identities

$$\begin{cases} \gamma^\nu \gamma^\mu \gamma_\nu = -2\gamma^\mu \\ \gamma^\mu l^\mu \gamma^\mu l = 2l^\mu l^\nu \gamma_\nu - \gamma^\mu \bar{l}^2 \end{cases} \quad (7.9)$$

and Feynman parametrization to obtain the vertex contribution

$$\delta\Gamma^\mu = \frac{\alpha}{2\pi} \int_0^1 dz(1-z) \int_{-\infty}^{+\infty} \frac{d^4\bar{l}}{(2\pi)^4} \frac{\left[-\frac{1}{2}\gamma^\mu \bar{l}^2 + P(z)m^2\gamma^\mu\right]}{[\bar{l}^2 - M^2]^3} f((k+z\bar{p})^2) \quad (7.10)$$

in which $M^2(z, \bar{p}^2) = (1-z^2)m^2 + z\mu^2$

After a variable change and introducing a dimensionless variable X, we obtain

$$\delta\Gamma^\mu = \frac{\alpha}{2\pi}\int_0^1 dz\,(1-z)\int_0^\infty dX\,\frac{X(X-1)}{[X+1]^3}f^2\left(X\frac{M^2(z,\bar{p}^2)}{\Lambda^2}\right) \quad (7.11)$$

We apply the method analog to the self-energy calculation, we then obtain

$$\delta\Gamma^\mu = \frac{\alpha}{4\pi}\left[\ln\left(\frac{\eta^2\Lambda^2}{m^2}\right) - 2\ln\left(\frac{\mu^2}{m^2}\right)\right] + \sigma\left(\frac{m^2}{\Lambda^2},\frac{\mu^2}{\Lambda^2}\right) \quad (7.12)$$

This result is equivalent to those obtained in others regularizations procedure. In our case the Ward-Takahashi identity was verified since

$$\delta\Gamma^\mu = -\frac{\partial}{\partial\bar{p}}\Sigma(\bar{p}^2)\Big|_{\gamma p=m} = \frac{\alpha}{4\pi}\left[\ln\left(\eta^2\frac{\Lambda^2}{m^2}\right) - 2\ln\left(\frac{\mu^2}{m^2}\right)\right] + \sigma\left(\frac{m^2}{\Lambda^2},\frac{\mu^2}{\Lambda^2}\right) \quad (7.13)$$

This equation proves that introducing Gaussian as test function reduced to partition of unity provide, at one loop order, self-energy and vertex satisfying the Ward-Takahashi identity.

## 8. Conclusion

It is shown that Gaussian functions may be used as test function in OPVD formulation of a Quantum Field Theory to deal with divergence. QFT is then defined, not in point space-time but smeared in an area described by the Gaussian test function. We see that quantum scalar Fields obey Klein-Gordon equations due to the properties of test functions. We obtain expressions of the vacuum fluctuation and Feynman propagators which contain Gaussian regularization factor. Loop calculation is performed in the case of one loop tadpole diagram and divergence-free result was obtained.

The Gaussian functions have some mathematical and physical properties which may make this approach interesting. The Heisenberg uncertainty principle is incorporated and it is represented by the factor $\mathcal{B}$. We have considered the case in which the mass m $\ll \sqrt{\mathcal{B}}$. This case yields finite result and we remark that the amplitude is related to the value $\mathcal{B}$ of the dispersion (statistical variance) of the energy and components of spatial momentum. The standard QFT correspond to a large value of $\mathcal{B}$. The main advantage of this formulation is that it needs no counter-parts to treat divergences in the higher order in perturbation theory. The dimension of space-time is not modified like in the dimensional regularization.

In the second part of the work, we reduce the test function in partition of unity. After applying the Taylor formula in the singular distribution, we obtain finite amplitude. Gauge and Poincaré invariance is kept at one loop level and it was verified by calculating the Ward-Takahashi identity.